# Fast Vascular Ultrasound Imaging with Enhanced Spatial Resolution and Background Rejection

Avinoam Bar-Zion, Charles Tremblay-Darveau, Oren Solomon, Dan Adam and Yonina C. Eldar

*Abstract*— **Ultrasound super-localization microscopy techniques presented in the last few years enable non-invasive imaging of vascular structures at the capillary level by tracking the flow of ultrasound contrast agents (gas microbubbles). However, these techniques are currently limited by low temporal resolution and long acquisition times. Super-resolution optical fluctuation imaging (SOFI) is a fluorescence microscopy technique enabling sub-diffraction limit imaging with high temporal resolution by calculating high order statistics of the fluctuating optical signal. The aim of this work is to achieve fast acoustic imaging with enhanced resolution by applying the tools used in SOFI to contrast-enhance ultrasound (CEUS) plane-wave scans. The proposed method was tested using numerical simulations and evaluated using two in-vivo rabbit models: scans of healthy kidneys and VX-2 tumor xenografts. Improved spatial resolution was observed with a reduction of up to 50% in the full width half max of the point spread function. In addition, substantial reduction in the background level was achieved compared to standard mean amplitude persistence images, revealing small vascular structures within tumors. The scan duration of the proposed method is less than a second while current super-localization techniques require acquisition duration of several minutes. As a result, the proposed technique may be used to obtain scans with enhanced spatial resolution and high temporal resolution, facilitating flow-dynamics monitoring. Our method can also be applied during a breath-hold, reducing the sensitivity to motion artifacts.**

*Index Terms*— **Contrast enhanced ultrasound; High order statistics; Super-localization microscopy; Super-resolution; Super-resolution optical fluctuation imaging**

## I. Introduction

Contrast-enhanced ultrasound (CEUS) imaging uses gas microbubbles encapsulated by phospholipid or protein shells as ultrasound contrast agents (UCAs). These contrast agents are injected into the blood stream. Since they are similar in size to red blood cells they are purely intravascular and flow throughout the vascular system. CEUS can be used for imaging of perfusion at the capillary level [1],[2] and for estimating blood velocity in small vessels (arterioles) by applying Doppler processing [3], [4]. However, the resolution of these scans is limited by the point spread function (PSF) of the system, reducing the clinical usability of CEUS. The goal of this work is to enable enhanced spatial resolution with clinically relevant acquisition times, while reducing the background noise level. To achieve this, we adapted a super resolution technique for fluorescence microscopy, specifically tailored to the characteristics of high concentration CEUS scans.

The mean radius of UCA microbubbles ranges between 1-3μm [5], [6]. However, they emit significantly broader echoes due to the PSF of the system. The size of the ultrasound scanner resolution cell is defined by the −6dB point of the beam profile in the axial, lateral and out-of-plane directions. Axial resolution of classic ultrasound scans $R_A$ is limited by the wavelength of the transmitted pulse and ranges between 200μm and 1mm depending on the insonation frequency [7] :

$$R_A = \frac{c\tau}{2}, \quad (1)$$

where c is the speed of sound in tissue and $\tau$ is the pulse duration. Lateral resolution is typically worse and depends on the transducer aperture and the imaging depth as well as the beamforming algorithm used. With focused beams [8]:

$$R_L = 1.2\frac{\lambda z}{L}, \quad (2)$$

where z is the depth and L is the transducer aperture. Out-of-plane resolution is lowest since a single out-of-plane focus point is typically set for the entire scan. In addition, since the SNR of a scan is depth dependent and CEUS signals are relatively weak, small vessels are hard to differentiate from the background. These limitations prevent classic CEUS from resolving the full structure and morphology of the vasculature.

Similar to ultrasound imaging, spatial resolution limitations are of great concern in optical fluorescent microscopy. Techniques such as PALM [9] and STORM [10] which gained increasing popularity in the last decade, have revolutionized fluorescence microscopy by enabling a ten-fold improvement in spatial resolution using precise localization of photo-

This project has received funding from the European Union's Horizon 2020 research and innovation program under grant agreement No. 646804-ERC-COG-BNYQ.

A. Bar-Zion (e-mail: barz@tx.technion.ac.il) and D. Adam (e-mail: dan@biomed.technion.ac.il) are with the Department of Biomedical Engineering, Technion—Israel Institute of Technology, Haifa 32000,

C. Tremblay-Darveau is with the Department of Medical Biophysics, University of Toronto, Toronto, ON, Canada, M5G1L7 (e-mail: charles.tremblay.darveau@mail.utoronto.ca).

O. Solomon (e-mail: orensol@tx.technion.ac.il) and Y. C. Eldar (e-mail: yonina@ee.technion.ac.il ) are with the Department of Electrical Engineering, Technion—Israel Institute of Technology, Haifa 32000.

switchable fluorophores. Inspired by super localization techniques in fluorescence microscopy, different approaches for achieving super-localization in CEUS were proposed in [11], [12], [13]. These methods are commonly referred to as ultrasound localization microscopy (ULM). By identifying resolvable echoes emitted from single microbubbles and fitting them with the transducer PSF, each microbubble is localized to the center of the respective PSF resulting in sub-wavelength localization of the microbubbles. In order to apply this processing successfully and reduce the probability of two microbubbles residing in the same resolution cell, the distribution of the microbubbles in the vasculature has to be sparse. As CEUS scans frequently include both large vessels with high number of bubbles flowing closely in the same streamline and small vessels where single microbubbles can be observed, this requirement has been met by using low concentration of UCA [12], [13] or bursting sub-populations of microbubbles [11].

Recently, a new method called ultrafast ultrasound localization microscopy (uULM) succeeded in producing high spatial resolution CEUS images using high UCA concentrations and low mechanical index, while maintaining the integrity of the bubbles. By using spatio-temporal filtering, the CEUS movies were decomposed to images containing sparse distributions of microbubbles [14] suitable for super-localization detection. These microbubbles were localized and tracked using ultra-fast ultrasound imaging which image the entire imaging plane at once by transmitting ultrasonic plane waves, achieving frame rates of 5kHz and more [15]. Using uULM, *in-vivo* microscopic images with penetration depth of a few centimeters and improved acquisition time were achieved. Currently, however, long acquisition times and low temporal resolution are still the main limitation of all ultrasound super-localization methods.

To improve spatial resolution in optics with better temporal resolution than PALM and STORM, the authors of [16], [17] proposed a method referred to as super-resolution optical fluctuation imaging (SOFI). This improvement is achieved by allowing for high emitter density which enables faster imaging. The emitters are overlapping in time, limiting the use of localization techniques, but contain information in the temporal fluctuations of each pixel throughout the movie. SOFI exploits temporal statistics of blinking emitters, registered throughout the captured movie in the form of cumulants [16], to effectively diminish the PSF width and reject out-of-focus light. However, when applying high-order statistics, the dynamic range of the reconstructed image increases and the imaged structures appear grainy. SOFI images have lower spatial resolution compared to super-localization techniques but their temporal resolution is significantly higher.

Frame to frame changes in CEUS scans are analogous to the statistical blinking of fluorescence sources in microscopy movies. In this paper we exploit this analogy and investigate the use of SOFI for CEUS signals. There are, however, a few important differences between the two signals. First, fluorophores are static and truly blinking. In contrast, when imaging with a low mechanical index, changes between CEUS frames are mainly the result of microbubbles carried by the blood flow within different blood vessels. These bubbles change their position relative to the transducer and are carried in and out of the imaging plane, causing decorrelation of the received RF single pixel time-series with a time-scale of milliseconds [14], resulting in an effective blinking. The flow of microbubbles along the vessels results in correlated fluctuations in neighboring pixels located along the stream lines and uncorrelated fluctuations in different vessels. As a result, we obtain a non-isotropic effect: improved separation between neighboring blood vessels and a smooth depiction of the blood vessels along the stream-lines. Moreover, the movement of the microbubbles enables clear depiction of vessel segments from single bubbles. Second, in fluorescence microscopy the diffraction limited image is a superposition of non-coherent light sources. In contrast CEUS signals in each frame are coherent. Nevertheless, superposition can be assumed in the RF/IQ signals rather than in their envelopes. Therefore, in CEUS we apply SOFI to the RF/IQ signal. Finally, CEUS signals emitted from large vessels with high UCA concentration follow a Gaussian distribution, rendering all cumulants of order high than two to zero. Instead of cumulants we therefore use moments of the IQ signals.

In this work, the statistical tools used in SOFI are adapted and applied to CEUS signals in order to produce fast acoustic scans with improved spatial resolution and reduced background noise level. The proposed technique uses the relative advantages of CEUS imaging producing non-invasive, high resolution scans of tumor and kidney vasculature at a depth of several centimeters. The local UCA concentration depends on the blood vessel size: in the main blood vessels, we can find several microbubbles per resolution cell, while single microbubbles are detected in smaller peripheral vessels. Therefore, we focus on CEUS imaging with high UCA concentrations that enable us to increase the portions of the vascular system sampled using short acquisition times. The effect of local UCA concentration on the resolution will be discussed in detail in the following sections.

The importance of achieving short acquisition times and good temporal resolution using methods like SOFI is two-fold. First, there are applications in which fast changes in blood flow inside the microvasculature can reveal important physiological information. One such case is imaging of neural activity based on changes in the blood flow [18]. Another example is imaging of blood flow in tumors in which fluctuations in blood flow are known to cause hypoxia and re-oxygenation, affecting tumor progression and response to treatment [19], [20]. Second, long acquisition times increase the effect of motion artifacts that may include out-of-plane motions and imaging plane shifts. While the transducer movement can be limited by mechanical mounting, breathing motion remains a fundamental limitation.

The rest of the paper is organized as follows: In Section II, a parametric model linking the CEUS signal to the underlying vasculature is introduced in order to facilitate analysis of the expected resolution gain. Our technique was validated using numerical simulations and *in-vivo* scans that are described in Section III. In Section IV, we show improvement in spatial resolution and background removal. Our results are discussed and analyzed in Section V. Section VI concludes the paper.

## II. THEORY

In this section we introduce a parametric model for CEUS signals and use it to explain how high order statistics of these signals provide high-resolution images of the underlying vasculature. These statistical measures can be used as a post-processing technique for improving the quality of displayed CEUS images.

### A. Clutter removal in CEUS imaging

A typical CEUS imaging cycle is composed of transmission of short ultrasonic pulses and reception of echoes reflected from the microbubbles and the surrounding tissue. The signals received in each transducer element are combined to produce a focused image of the scanned region in a process called beamforming. Following demodulation, the complex analytic CEUS signal may be written as

$$f(x,z,t) = I(x,z,t) + iQ(x,z,t), \quad (3)$$

where $I(x,z,t)$ is the in-phase signal component and $Q(x,z,t)$ is the quadrature signal component. The received signal $f(x,z,t)$ is a superposition of echoes originating from the microbubbles carried by the blood flow and the strong clutter signal originating from the surrounding tissue. Following [21], we model the received IQ signal as

$$f(x,z,t) = c(x,z,t) + b(x,z,t) + n(x,z,t), \quad (4)$$

where $c(x,z,t)$ is the clutter signal reflected from the tissue, $b(x,z,y)$ is the CEUS signal produced by the microbubbles and $n(x,z,t)$ is additive noise.

At low acoustic pressures (e.g. low mechanical indices) microbubbles produce non-linear (harmonic) echoes compared to the mostly linear tissue scattering. Different pulse streams and processing algorithms were designed to make use of this physical phenomenon and remove tissue related background from CEUS signals [22]. In addition, different priors on the relatively slow movement of the tissue [23] and its spatial coherence [21] are used in order to design temporal and spatio-temporal (SVD based) filters for tissue clutter removal. Using FIR/IIR filters as in [23], the clutter free movie $\hat{b}(x,z,t)$ is estimated by removing the quasi static tissue signal. In our derivations below, we assume that $\hat{b}(x,z,t)$ is a good estimate of the clutter free signal so that subsequent processing is performed on this estimate. For convenience, we use $b$ instead of $\hat{b}$ to represent the estimated UCA signal.

### B. Parametric model for CEUS signals

In order to introduce our processing approach we first present a model for the CEUS signal. This signal includes a variety of UCA concentrations: large blood vessels may contain many microbubbles in each resolution cell while poorly perfused areas such as tumor vessels might consist of a sparse distribution of bubbles. The effects of different UCA concentrations on the reconstruction quality of our method will be discussed in detail in Section V. In general, for each time $t$, the signal measured at every point $r = (x,z)$ (lateral and axial directions, respectively) is influenced by a time dependent set of microbubbles $K(t)$ located at time dependent positions $r_k(t) = (x_k(t), z_k(t))$. That is, $K(t)$ is an index set corresponding to the in-plane bubbles for specific $t$. The overall signal at a vessel segment positioned at $r = (x,z)$ is then given by

$$b(x,z,t) = \sum_{k \in K(t)} H(x - x_k(t), z - z_k(t)) \cdot \sigma_k, \quad (5)$$

where $H(x,y)$ is the system's PSF and $\sigma_k$ is the scattering cross section of each microbubble. The PSF of the scanner substantially blurs the resulting CEUS images, limiting the detection and separation of small blood vessels and the analysis of fine morphological details (Fig. 1a).

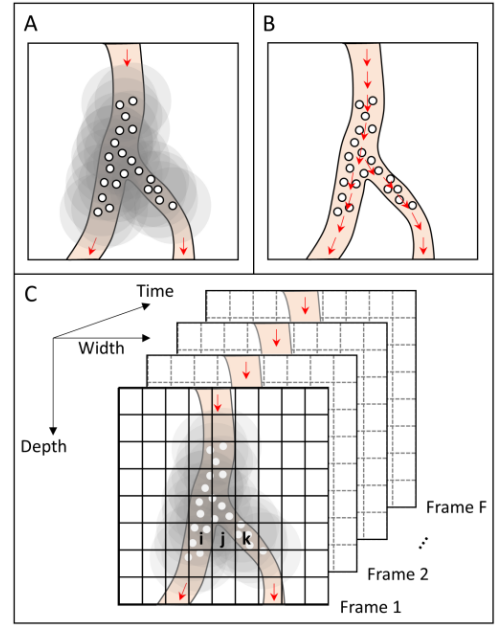

Fig. 1. Principles of CEUS fluctuation imaging. (A) The signal measured at a given position in the vasculature is influenced by $K$ close-by microbubbles. The echo from each microbubble is blurred by the system PSF. (B) Microbubbles are carried by the blood flow, changing their relative position at each frame. (C) The locations of the microbubbles are discretized and associated with one of $N$ neighboring pixels. As a result of the blurring effect, pixels can have a non-zero value even if they do not include any microbubbles. This blurring may be reduced by calculating the time-series statistics in each pixel, effectively shrinking the PSF width.

The microbubbles are carried by the blood flow changing their relative locations and producing a fluctuating CEUS signal. Microbubbles flowing in different streamlines move independently while microbubbles flowing in the same streamline move in a correlated fashion (Fig. 1b, see assumptions below). The locations of the bubbles can be associated with one of $N$ neighboring volume cells in which they are located at each time point (Fig. 1c): By dividing the CEUS images to pixels of size $[\Delta x, \Delta z]$, we formulate (3) as,

$$\begin{aligned} b(x,z,t) &= \sum_{k \in K(t)} H(x - x_k(t), z - z_k(t)) \cdot \sigma_k \\ &= \sum_{n=1}^{N} \sum_{k \in K(t)} H(x - x_k(t), z - z_k(t)) \cdot \\ &\quad \sigma_k \cdot 1_{\substack{|x - x_n| \le \Delta x/2 \\ |z - z_n| \le \Delta z/2}} (x_k(t), z_k(t)), \end{aligned} \quad (6)$$

where $r_n = (x_n, z_n)$ is the location of the n$^{th}$ volume cell, and $1_P(x, z)$ is an indicator function, which admits the value 1 if $(x, z) \in P$ and 0 if $(x, z) \notin P$.

Next, we approximate (4) by discretizing the location of the bubbles according to the volume cell in which they are located and looking at the fluctuations in each volume cell instead of looking at single bubbles:

$$b(x,z,t) \approx \sum_{n=1}^{N} H(x - x_n, z - z_n) \cdot$$

$$\sum_{k \in K(t)} \sigma_k \cdot 1_{\substack{|x-x_n| \leq \Delta x/2 \\ |z-z_n| \leq \Delta z/2}}(x_k(t), z_k(t)) \quad (7)$$

$$= \sum_{n=1}^{N} H(x - x_n, z - z_n) \cdot s_n(t),$$

where $s_n(t)$ is the time-dependent fluctuation in every volume cell, represented in the scan as a pixel. At a given time, each volume cell can include several microbubbles or no bubbles at all, depending on the structure of the vasculature and the blood flow patterns. Using this notation, at a specific time the value of $s_n(t)$ may be negligible even if many microbubbles are located in the surrounding pixels due to destructive interference of the echoes from different microbubbles.

To proceed, we rely on the following assumptions:
1. For each processed ensemble, $s_n(t)$ is a wide sense stationary process. This assumption is reasonable if the average concentration of the microbubbles in each volume cell does not change during the processed time period and the tissue is static. Furthermore, stationarity is implicitly assumed by all super-localization methods estimating blood velocity [12], [14]. This assumption is more challenging when long acquisition times are used.
2. In volume cells located in distinct blood vessels various microbubbles are flowing with different velocities and directions. Assuming that $j$ and $n$ are volume cells located at different blood vessels, the fluctuations $s_j(t)$ and $s_n(t)$ are statistically independent. A direct consequence is statistically uncorrelated fluctuations, or $E\{\tilde{s}_j(t_1)\tilde{s}_n(t_2)\} = 0, \forall j \neq n$ where $\tilde{s}_n(t) = s_n(t) - E\{s_n(t)\}$.

It is important to note that in peripheral flow, CEUS signals originating from volume cells that belong to the same streamline are correlated since the bubbles flow from one such volume cell to the other. The independence of bubbles in neighboring volume cells and the dependence along the stream line will result in a non-isotropic effect when applying SOFI to CEUS that will be discussed in the next sub-section.

In small vessels with slow velocity and low UCA concentration a long observation time is needed to ensure the detection of at least a single bubble during acquisition time. Therefore, the limitation of short acquisition times is partial observation of the flow in small vessels. This is a physical limitation that is not method related and can be accommodated for by using high UCA concentrations.

*C. Application of SOFI to CEUS signals*

To apply SOFI to CEUS we calculate the time-dependent statistics of the CEUS signal in each pixel. This leads to an improvement in the spatial resolution of the scan by effectively diminishing the size of the PSF (Fig. 1c).

The autocorrelation function of the signals at a point $r$ is defined as

$$G_2(r, t_1, t_2) = E\{\tilde{b}(r, t_1) \cdot \overline{\tilde{b}(r, t_2)}\}, \quad (8)$$

where $\tilde{b}(r, t) = b(r, t) - E\{b(r, t)\}$ and $b(r, t)$ is given by (5). Using assumption 1, the autocorrelation function at a point $r$, can be written as

$$G_2(r, \tau) = E\{\tilde{b}(r, t) \cdot \overline{\tilde{b}(r, t + \tau)}\}$$

$$= \sum_{j,n} H(r - r_j) \cdot \overline{H}(r - r_n) \cdot E\{\tilde{s}_j(t + \tau) \overline{\tilde{s}_n(t)}\}, \quad (9)$$

where $j$ and $n$ run over all the volume cells in the image. Here, $\tau$ represents the time lag used in the evaluation of the correlation function. From assumption 2, the cross terms of independent volume cells in (7) are zero, and (7) can be written as

$$G_2(r, \tau) = \sum_{n} |H|^2(r - r_n) \cdot E\{\tilde{s}_n(t + \tau) \overline{\tilde{s}_n(t)}\}$$

$$+ \sum_{\substack{i,l \\ i \neq l}} H(r - r_i) \cdot \overline{H}(r - r_l) \cdot E\{\tilde{s}_i(t + \tau) \overline{\tilde{s}_l(t)}\}, \quad (10)$$

where $i$ and $l$ are indices of (dependent) volume cells located in the same stream line and $n$ runs over all volume cells in the image.

For different blood vessels, the cross terms in (10) equal zero and the autocorrelation function can be written as a sum of squared PSFs leading to a sharper image compared to the original acquired frames. Approximating the PSF as a 2D Gaussian function, the resulting spatial resolution is improved by approximately a factor of $\sqrt{2}$ [16]. CEUS signals originating from different volume cells that belong to the same streamline are statistically dependent and therefore produce cross terms that include the first order PSF. However the resulting smoothing effect in the direction of the blood vessels does not reduce the quality of these images, since only separation between adjacent blood vessels is of interest. A simulation showing this non-isotropic effect is presented in section III.

Assuming that over a short time period the signals are stationary and ergodic, the expected value in (10) is estimated using temporal averaging for every value of $\tau$. The described technique produces a single estimation of the autocorrelation for each pixel of processed CEUS movie. Displayed together as an image (for specific prechosen $\tau$), these estimations create an image of the underlying vasculature with improved spatial resolution. It is important to note that the high background level in CEUS imaging is mainly the result of thermal noise manifested as additive white Gaussian noise. The autocorrelation of this noise is a delta function in time and space. Therefore, improved SNR may be achieved by using $\tau > 0$. Moreover, in peripheral vessels the decorrelation time of single pixel CEUS time-series is on the order of a few milliseconds, due to the limited blood velocity range. Using plane wave imaging with a PRF larger than 5kHz, echoes from the same bubble register in several consecutive frames. Hence, correlations can be calculated even for $\tau > 0$.

In the spirit of SOFI it is possible to utilize higher order statistical computations to achieve additional improvement in spatial resolution. This can be realized by calculating higher-order statistics (HOS) in terms of cumulants. Consider the case of a pixel which is affected by two bubbles located in different blood vessels, implying that the cross-term in (10) equals zero. Calculation of the $p^{th}$ moment of a given pixel may yield additional terms with the PSF raised to a power smaller than $p$ [16]. Such broader PSF will obscure the desired narrow PSFs to the power of $p$, limiting resolution enhancement. In cumulants, only the PSF to the power of $p$ is maintained, and resolution enhancement is attained due to the narrowing of the PSF, even for $\tau \neq 0$.

### D. CEUS imaging: moments versus cumulants

Following [24], the cumulants (also called semi-invariants) of a given random vector $x \in \mathbb{R}^m$ are defined through the cumulant generating function

$$\Psi(v) = \log E\left\{e^{jv^T x}\right\}, \quad (11)$$

with $v \in \mathbb{R}^m$. The cumulant of order $p$ is the $p^{th}$ coefficient of the Taylor series expansion of $\Psi(v)$ around 0. Since (11) is the log of the moment generating function, the cumulant of order $p$ is defined in terms of its joint moments up to that order. For further details about cumulants we refer the reader to the tutorial [24], which provides detailed explanations on the calculations of cumulants in terms of moments.

Auto-cumulants (AC) are cumulants which are computed using time traces of the same pixel. The calculation is performed for each pixel's time-trace independently. The $p$'th order AC of (5) is given by,

$$AC_{p,b}(r, \tau_1, ..., \tau_{p-1}) = \mathrm{cum}\left\{\sum_{n=1}^{N} H(r-r_n)s_n(t), ..., \sum_{n=1}^{N} H(r-r_n)s_n(t+\tau_{p-1})\right\}. \quad (12)$$

Under the assumption of statistically independent fluctuations, the latter equation may be written as

$$AC_{p,b}(r, \tau_1, ..., \tau_{p-1}) = \sum_{n=1}^{N}|H|^p(r-r_n)AC_{p,s_n}(\tau_1, ..., \tau_{p-1}). \quad (13)$$

Different time-lags produce different images, though it is common in optics to choose all time-lags as zero. It is evident from (13) that $AC_{p,b}$ contains only the $p^{th}$ power of the PSF, yielding a theoretical resolution enhancement of factor $\sqrt{p}$.

Despite the high motivation for using cumulants instead of moments to achieve resolution enhancement, [25] showed that the distribution of the received CEUS IQ signal for large blood vessels follows a Gaussian distribution. This renders all cumulants higher than 2 to be equal zero. Therefore, we applied high order moments to the complex signal and high order cumulants to the envelopes of the received signal, where the even order symmetric $p^{th}$ order moment of a complex random variable is defined as [26]:

$$G_p(r,\tau) = E\left\{\left(\tilde{b}(r,t)\cdot\overline{\tilde{b}(r,t+\tau)}\right)^{p/2}\right\}. \quad (14)$$

In the results section we perform a detailed examination of the two approaches and point out some differences between them. The *in-vivo* results presented in Section IV suggest that using moments is preferable for CEUS signals over the cumulant based approach.

Traditionally, the envelopes of the ultrasound signals are used for display:

$$A(x,z,t) = \sqrt{I(x,z,t)^2 + Q(x,z,t)^2}. \quad (15)$$

It is interesting to note that when the mean of the IQ signal is negligible, the zero time lag central moments (zero-mean moments) calculated from the IQ signals are equivalent to the (non-central) moments calculated from the ultrasound signal envelope. For example, the second moment can be expressed as:

$$G_2(x,z,\tau=0) = E\left\{\tilde{b}(x,y,t)\cdot\overline{\tilde{b}(x,y,t)}\right\}$$
$$= E\left\{I(x,z,t)^2 + Q(x,z,t)^2\right\} = E\left\{A(x,z,t)^2\right\}. \quad (16)$$

In these cases, the desired even order moments may be derived directly from the filtered linear envelope data. This however, cannot be done for low UCA concentration.

In this work the moments and auto-cumulants were calculated for $\tau = 1$ (time difference between two consecutive frames), significantly reducing the background level. High order statistical measures were calculated from the movie time-traces of each pixel. Following the theoretical analysis presented in this section, the proposed technique is evaluated using numerical simulations and *in-vivo* scans in the next sections.

## III. MATERIALS AND METHODS

### A. Simulations

In this section we performed two types of simulations. First, a series of simulation tests were performed to characterize the ability of moments to substitute for cumulants when aiming to achieve enhanced resolution in IQ ultrasound images. The simulations mimicked a cross-section of two vessels, located close to each other, which cannot be separated in the envelope image. The signal was simulated as:

$$b_{sim}(x,t) = a_1(t)\exp\left(\frac{-(x-\mu_1)^2}{\sigma^2}\right)\exp(-j\theta_1(t))$$
$$+ a_2(t)\exp\left(\frac{-(x-\mu_2)^2}{\sigma^2}\right)\exp(-j\theta_2(t)). \quad (17)$$

The following random variables were generated for each frame: $a_1(t)$ and $a_2(t)$ are independent random variables with Bernoulli distribution and success probability P, varying between different experiments and producing the flickering statistics; $\theta_1 \sim U(-\pi, \pi)$ and $\theta_2 \sim U(-\pi, \pi)$ simulate the random phase arising from the off-grid location of the bubbles. The centers of the two vessels are located at $\mu_1$ and $\mu_2$ and the Gaussian PSF has a STD of $\sigma$. The spacing between the two vessels was empirically selected to be $\mu_2 - \mu_1 = 1.1\sigma$ in order to produce an overlap between the envelopes. The experiment included 500,000 simulations for every success probability P = 0.4, 0.2, 0.1, 0.05, 0.025, 0.0125. These values cover a wide range of microbubble density relevant for small vessels like those imaged in the in-vivo experiments. The second and fourth moments were calculated from the simulated complex signals (according to (14)), together with the temporal mean of the signal envelope. The sum of the fourth order envelopes of the two vessels served as a reference for ideal resolution improvement.

In the second simulation, we further investigated the anisotropic smoothing effect of applying statistical calculations to CEUS signals. We have performed a simulation of UCA flow within two streamlines. Three Gaussian bubbles were simulated in the upper vessel and six in the lower vessel. All the dimensions in the simulation were normalized to the standard deviation (std) of these Gaussians. A total of 150 frames were generated. The two streamlines are separated by $1.1 \cdot \sigma$. The flow velocity in the upper vessel is $0.25 \cdot \sigma/frame$ while the flow velocity in the lower vessel is $0.5 \cdot \sigma/frame$. Similar to the first simulation, the second and fourth moments were calculated from the simulated signals, together with the temporal mean of the signal envelope.

*B. In-vivo Ultrafast Imaging*

The proposed algorithm was evaluated using two New Zealand white rabbit models: normal vasculature was imaged in the kidneys of healthy rabbits using pulse inversion Doppler [27] and a VX-2 tumor xenograph was imaged using amplitude-modulation-pulse-inversion Doppler (AMPID, 3 pulse sequence) [4]. The rabbits weighed 3.5-4.5kg. For tumor imaging, approximately $10^6$ VX-2 carcinoma cells were injected intramuscularly in the hind limb of a rabbit. The tumor was monitored until reaching a diameter of over 1cm. The *in-vivo* scans were performed using an Aixplorer diagnostic ultrasound system (Supersonic Imagine, Aix-en-Provence, France) with an 8MHz probe (L15-4 linear array). Only part of aperture was used upon reception due to channel limitation. Single cycle pulses with a carrier frequency of 4.5MHz were used in order to capture the harmonic components. Entire images were acquired using plane-wave acquisitions at an imaging frame rate of 5/3 kHz (PRF of 5 kHz) and low mechanical index of 0.06. Short cine loops of 150 and 1000 frames were acquired for the kidney and tumor models, respectively without using multi- angle compounding. Definity (Lantheus Medical Imaging Inc., N. Billerica, MA, USA), a clinically approved commercial ultrasound contrast agent (UCA) with lipid shell, was used in all experiments. The microbubbles were injected via the ear vein of the rabbit in bolus injections of 0.5mL with concentration of 10 µL/Kg and flushed with an additional 1mL of saline. All the experimental protocols were approved by the Sunnybrook Research Institutional review board.

*C. Radio-frequency and Wall Filtering*

The RF echoes were first summed in order to segment the signal nonlinear component from the linear clutter. The IQ signal was demodulated by the scanner via quadrature sampling, beamformed with a delay-and-sum algorithm and filtered by an FIR bandpass filter with a 2 to 12MHz pass-band. A Doppler filter (i.e. wall-filter) was used to separate the echoes of flowing microbubbles from the weak harmonic tissue clutter and stationary microbubbles. Following [3] and [23], a sixth order high pass Butterworth filter with a projection initialization was used. The cut-off frequency was chosen as 0.03·PRF and used for both kidney and tumor models. This is the lowest cut-off possible in order to include slow flowing microbubbles while removing clutter artifacts.

*D. Calculation of High-order Statistics*

Following application of the high-pass clutter filters and in the spirit of SOFI, high-order statistical analysis of single-pixel CEUS time-traces was performed. In this work, the temporal-mean of the envelope signal, the $2^{nd}$ and $4^{th}$ order central moments of the IQ signal (according to (12)) and the $4^{th}$-order cumulants of the envelope were calculated ($AC_4 = G_4 - 3G_2^2$ for the same $r$, $\tau$ [24]) for a single frame time-lag ($\tau = 1$). The spatial map of the calculated moments and cumulants produced high resolution images of the underlying vasculature.

IV. RESULTS

*A. Simulations*

We performed noiseless simulations of 1D cross-section of two close-by vessels with a CEUS signal fluctuating randomly over time, to assess the possible resolution improvement of different statistical orders. The fluctuation rates simulated different probabilities of microbubble appearing in each of the vessels at a given time. Signals originating from these two vessels are assumed to be statistically independent, and emit a signal with probability P, drawn from a Bernoulli distribution for each RF vessel. For all values of P (duty-cycle) used in this analysis, the separation between vessels achieved using the $4^{th}$ order moment was better than that achieved using the $2^{nd}$ order moment (Fig. 2). Furthermore, for low values of P the result was close to the ideal result expected from combining the $4^{th}$ order PSFs (Fig. 2). The steepness of the outer profile of the two vessels increased with the statistics order and was identical for the $4^{th}$ moment and the ideal $4^{th}$ order profiles.

These results are quantified in table 1 displaying the amplitude measured at the mid-point between the two vessels normalized to maximum value of each profile. The $2^{nd}$ moment and the ideal $4^{th}$ order values remain stable around 0.97 and 0.59 respectively. The mean envelope value is equal to 1. The $4^{th}$ moment value is 0.9 for P = 0.4 and decreases considerably for lower values of P until converging to the value of the ideal $4^{th}$ order plot.

Fig. 3a shows a single frame out of total of 150 frames of the 2D simulation. Fig. 3b shows the temporal mean of the movie. Clearly, no separation between the streamlines is achieved. In

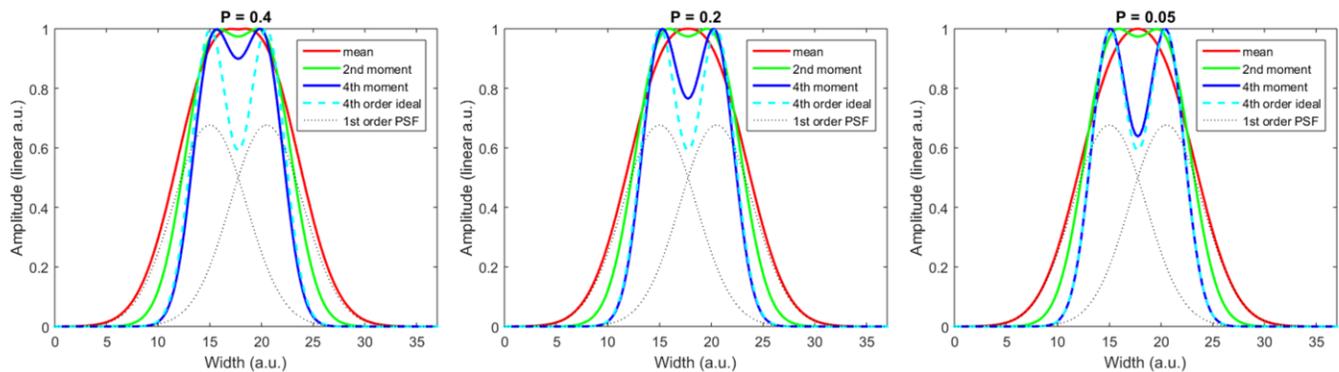

Fig. 2. Numerical simulations of vessel separation. Simulations of a 1D cross-section through two close-by vessels were performed for different probabilities of microbubbles flowing in any one of them at each time point (Bernoulli distribution with different values of P). The 4$^{th}$ order moment showed improvement in resolution compared to the 2$^{nd}$ moment profiles and got closer to the ideal 4$^{th}$ order profile for lower values of P.

Fig. 3c and Fig. 3d, the second and forth order moments are calculated. It is evident that as the statistical order increases, the separation between the two simulated blood vessels becomes more evident, while retaining good depiction of the streamlines themselves. Hence we achieve good separation between the vessels while preserving their overall shape, due to the anisotropic smoothing effect.

### B. In-vivo Experiments

We first demonstrated the proposed method by processing a short CEUS scan (150 frames) of a rabbit kidney showing normal and hierarchically ordered vasculature (Fig. 4). The different parts of the kidney's vasculature, observed in the unprocessed image, included a variety of sparsity levels: the main vessels are characterized by speckle patterns (Fig. 4a arrowheads) while single bubbles can be seen flowing in the smaller outer vessels (Fig. 4a arrows). The microbubbles flowing in small vessels can be matched and tracked over consecutive frames without additional processing. Time series taken from a wide variety of vessel sizes showed that while the signal in the main vessels fluctuates randomly due to the large number of microbubbles per resolution cell, in certain small vessels a single bubble was detected during the acquisition time (supplemental Fig. S1). This result matches a value of $P = 0.007$ in the simulations (1 bubble/150 frames).

To adjust for the intrinsic increase in the dynamic range of high order statistical images, the dynamic range of the log-transformed images was scaled to fit the order of the statistics: the dynamic range of images showing second-order statistics was doubled and the dynamic range of images showing 4$^{th}$ order statistics was quadrupled.

The visualization of small vessels in the kidney cortex improved in both the 2$^{nd}$ and the 4$^{th}$ moment images. Using high-order statistics, vessels that could not be resolved in the temporal-mean image (Fig. 4, dashed lines) were separated. The normalized amplitude profiles in these lateral and axial cross-sections are presented in Fig. 5a and Fig. 5b respectively. In the lateral direction two vessels that could not be resolved in the mean envelope plot were differentiated using the 2$^{nd}$ moment and separated using the 4$^{th}$ moment (Fig. 5a, around x = 33mm). In addition, separation between the two left vessels was improved using the high order moments (Fig. 5a, around x = 31.75mm). Similarly, a separation between two close vessels was observed in the axial direction (Fig. 5b). These two distinct vessels are observed in the original 4$^{th}$ order image (Fig. 4).

The resolution improvement was quantified by comparing the full-width-half-max (FWHM) of normalized amplitude profiles measured in lateral cross-sections through 5 small vessels (the specific locations are presented in supplemental Fig S2). The FWHM in the temporal mean image was 458.2±53.3 µm. An improvement in resolution was observed for both the second and fourth moments: the FWHM in the 2$^{nd}$ moment image was 316.1±25.1 µm (69.4±6.0 % of the temporal mean FWHM) while the FWHM in the 4$^{th}$ moment image was 227.3±9.0 µm (50.0±4.6 % of the temporal mean FWHM).

TABLE I
NORMALIZED VALUE AT MID-POINT BETWEEN VESSELS

| Probability P | Mean | 2$^{nd}$ Moment | 4$^{th}$ Moment | 4$^{th}$ Order Ideal |
|---|---|---|---|---|
| 0.4 | 1 | 0.97 | 0.9 | 0.59 |
| 0.2 | 1 | 0.97 | 0.76 | 0.59 |
| 0.1 | 1 | 0.97 | 0.69 | 0.59 |
| 0.05 | 1 | 0.97 | 0.64 | 0.59 |
| 0.025 | 1 | 0.97 | 0.61 | 0.59 |
| 0.0125 | 1 | 0.97 | 0.6 | 0.59 |

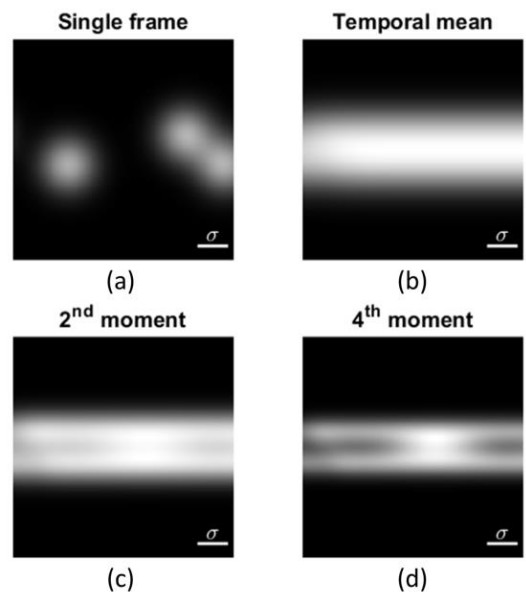

Fig.3. Simulation showing the anisotropic effect of SOFI applied to CEUS data. (a) Single frame from the simulated cine, from a total of 150 frames. (b) Temporal mean image. (c) Second moment image. (d) Fourth moment image. While high order statistics improve the separation between neighboring vessels the integrity of the vessels in the axial direction is not compromised.

In the second in-vivo experiment, the vasculature around and inside the hind-limb intramuscular VX-2 tumor was imaged (Fig. 6). The vessels inside the tumor were sparsely distributed and hence these vessels were resolvable against the poorly-perfused background. This scan, including 1000 frames, demonstrates the background rejection capabilities of the high-order statistics images. The mean envelope image of the tumor's vasculature showed high background level compared to the amplitude of the small vessels. The weak UCA signal and dominant background limited the effective dynamic range of the mean envelope image to around 23dB compared to around 30dB in the kidney scans. The background level in both the 4$^{th}$ moment and 4$^{th}$ cumulant images was dramatically lower compared with that of the mean image, allowing clear visualization of small vessel (dashed arrows). The 4$^{th}$ cumulant image showed zero-value regions in the lumen of some vessels (dotted arrow).

The ability of high order moments and cumulants to separate adjacent small vessels can be further studied by zooming on a bifurcation in the vasculature of the tumor that was not observed in the temporal mean envelope image but was resolved in the 4$^{th}$ moment image (Fig. 7). Although small, these vessels have a significant width compared to the PSF size. Applying statistical calculations helped separating the boundaries of the two vessels. Similar separation was observed in the 4$^{th}$ cumulant image. In addition, a contrast of over 60dB between the vessels and the background was observed in the 4$^{th}$ moment image, compared to 6dB in the mean envelope image.

Finally, the resolution improvement in the axial direction is estimated and quantified by comparing the normalized amplitude profile measured in a cross-section of a small vessel (Fig. 8). In this cross-section, the 4$^{th}$ moment produced an improvement of 40% (full-width-half-max) in the resolution compared with the mean envelope image (140.4 µm vs. 234.9 µm respectively). In the same cross-section the 2$^{nd}$ moment produced a resolution gain of 14% compared to the mean. Furthermore, this small vessel had low contrast in the mean envelope image that improved in the 4$^{th}$ moment image.

## V. DISCUSSION

In this work, CEUS scans of different vascular structures were processed in the spirit of SOFI by calculating the high-order statistics of single-pixel time-series. High UCA concentrations were used to achieve high temporal resolution while maximizing the portion of the vascular system imaged during short acquisition times. The proposed method was able to produce improved spatial resolution and background rejection compared to the mean amplitude persistence images. These capabilities were studied and quantified using simulations (Fig. 2 and Fig. 3) and *in-vivo* scans. In these scans we focused on two fundamental vascular structures: single resolvable vessels (Fig. 5 and Fig. 8) and bifurcations (Fig. 7).

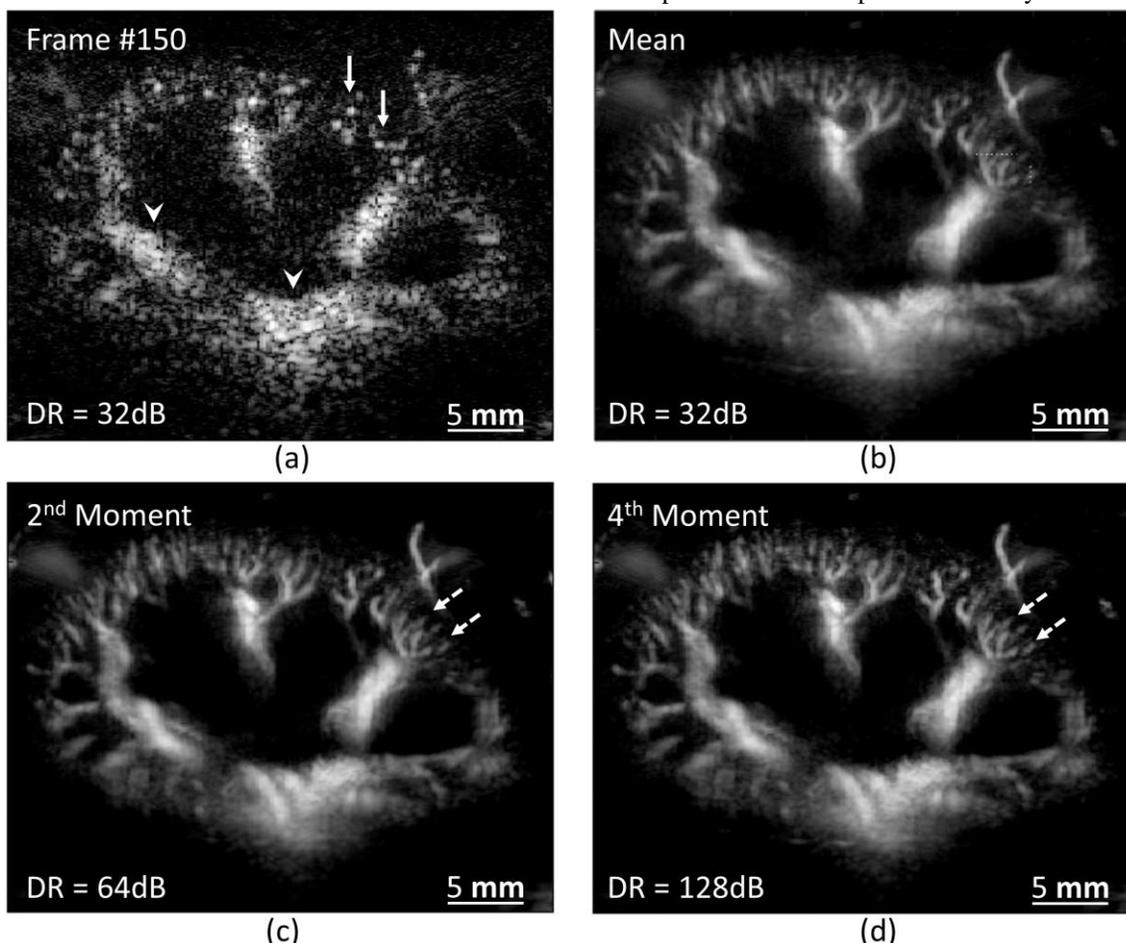

Fig. 4. Images of a healthy kidney. A single CEUS frame from the 150 frame long scan is compared to the mean envelope image and the 2$^{nd}$ and 4$^{th}$ order moments of the time series measured in each pixel. (a). The original scan shows different UCA concentration levels: speckle patterns are seen in large vessels (arrowheads) while single bubbles can be detected in the outer cortex vessels (arrows). Small vessels in the cortex of the kidney are resolved in the 2$^{nd}$ and 4$^{th}$ moment images (c and d respectively) but not in the mean envelope image (b, dashed arrows). Axial and lateral cross-sections through these vessels, further studied and analyzed in Fig 5, are marked by dashed lines.

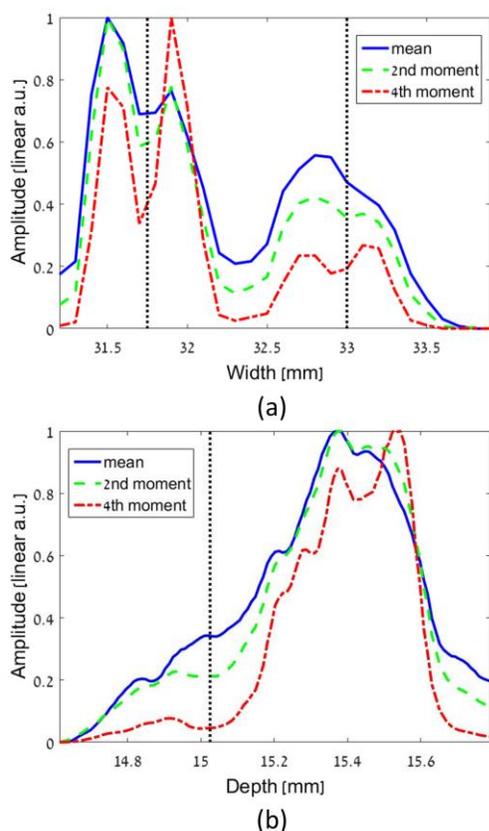

Fig. 5. Horizontal and vertical cross-sections in a kidney scan showing improved separation between close-by vessels. The horizontal (a) and vertical (b) cross-sections show improvement in the separation of close-by vessels using the 2$^{nd}$ and 4$^{th}$ moments, including vessels that were not resolved in the mean envelope image.

Despite the analogy between the blinking of the fluorophores in fluorescence microscopy and fluctuations in the CEUS signal, there are several main differences between these modalities that demanded the modification of SOFI when applied to CEUS signals. Due to the coherence of the ultrasonic signal, superposition of high order PSFs was shown using the (complex) IQ signal and not the envelope. In addition, when processing the CEUS IQ data, high order statistics in the form of moments and not cumulants were used in order to be compatible with the statistics of these signals. It is important to note that in contrast to florescence microscopy, CEUS signals measured in different pixels located along a streamline cannot be assumed to be independent. Therefore, simulations have shown that while different stream lines will be better resolved using high-order statistics, the integrity of each streamline will be preserved. This outcome is favorable in CEUS imaging since the structure of the vasculature and not the specific location of each microbubble is of interest.

Theoretically, cumulants should be used to achieve super-resolution in SOFI, as they fully remove the cross terms between different sources. Since the use of cumulants in CEUS IQ signals is limited to second order statistics, the use of high-order moments was investigated in this work. The simulations performed as part of this study showed that the 4$^{th}$ order moment can improve the separation of small vessels and obtains a separation similar to the ideal sum of 4$^{th}$ order PSFs for low concentrations. In addition, the 4$^{th}$ order moment of the IQ signal and the 4$^{th}$ order cumulant of the envelope were compared using a CEUS scan of a tumor (Fig. 6). The resulting images were similar. However, small zero-valued regions appeared in the lumen of some vessels in the cumulant image. Few small vessels that appear in all the other images could not be observed in the 4$^{th}$ order cumulant image. From these results, it seems that the moments of the IQ signal should be used when processing CEUS scans.

The *In-vivo* scans analyzed in this work included bolus injections which led to high UCA concentrations in large blood vessels and low concentrations in smaller vessels. The simulations have shown improvement in the spatial resolution using the 4$^{th}$ order moment compared to the temporal mean and 2$^{nd}$ order moment for a wide range of duty-cycles (concentrations). The separation achieved using the 4$^{th}$ order moment improved monotonically when reducing the duty-cycles. Therefore, the low UCA concentrations in small vessels could assist in their separation. It should be noted that reducing the concentration of injected UCA might reduce the influence of low order PSFs on the moments. However this approach necessitates longer acquisition times to ensure imaging of sporadic bubble flow within the capillaries.

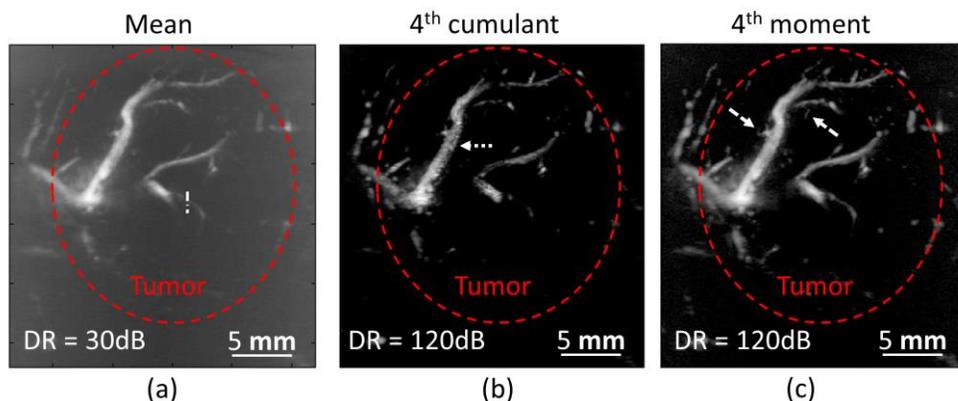

Fig. 6. Images of a VX-2 tumor. The mean envelope image is compared to the 4$^{th}$ order cumulants and moments of the time series measured in each pixel. (a) High background level was observed in the temporal-mean image. An axial cross-section through a small vessel used for estimating the resolution of each image as presented in Fig 8 is marked by a dashed line. The background reduction capabilities of the 4$^{th}$ order moments and cumulants (b and c respectively) enable a clear depiction of small vessels (dashed lines). Zero value areas in the lumen of some vessels are seen in the 4$^{th}$ order cumulants image (dotted lines).

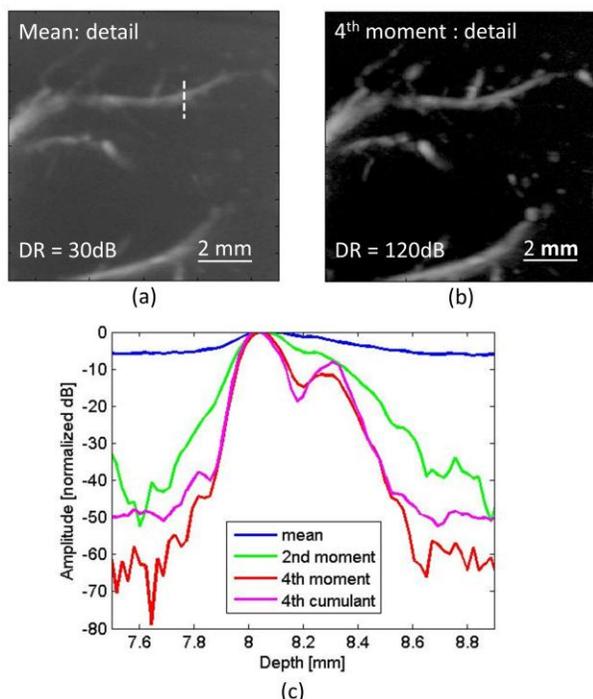

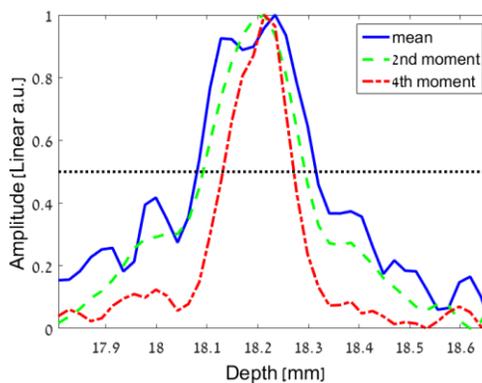

Fig. 7. Cross-section through a small bifurcation in the VX-2 tumor scan. The structure of the bifurcation (dashed line), blurred in the mean envelope image, is resolved in the 4$^{th}$ moment image. In addition, the background level is significantly reduced in the 4$^{th}$ moment image.

Fig. 8. FWHM obtained using the different statistics. The normalized amplitudes of the different images, measured along a cross-section in a small vessel (dashed line in fig, 6), are compared. A reduction in the FWHM is achieved using the 2$^{nd}$ and 4$^{th}$ moments.

Similar to [16], the theoretical changes in the spatial resolution of the images as a function of the calculated statistics order were analyzed using a model for CEUS signals. Noiseless simulations of adjacent blood vessels show an improvement in their separation, compared with the mean image, supporting our claim that high-order statistical calculations manage to reduce the PSF width. In addition, the resolution gain of the different statistics was measured by zooming-in on cross-sections in small vessels. The measured resolution gain of the 4$^{th}$ order moment in the lateral direction (50.0±4.6 %) was similar to the theoretical value (50%) while the measured resolution gain of the 4$^{th}$ order moment in the axial direction (40%) was lower than the theoretical value. This could be attributed to the finite width of the investigated vessels that were not compensated for in our analysis.

Generally speaking, good temporal resolution and short acquisition times are among the most important advantages of ultrasound imaging. In CEUS this attribute is used for tracking the dynamics of blood flow. Even during the time period in which the UCA are assumed to be uniformly mixed in the blood, the UCA concentration may change depending on local blood flow fluctuations. Imaging of neural activity in the brain and fluctuation of flow in tumors are two examples in which these changes are of great importance and good temporal resolution is crucial.

The statistical techniques used in this work facilitated CEUS imaging with improved spatial resolution and good temporal resolution. Using these methods, the entire CEUS scan can be divided to a series of partially overlapping ensembles that when analyzed separately and put together describe changes in blood flow. There is however a limitation for how short the processed ensembles may be. The precision in which the statistics can be estimated depends on the number of samples in the processed ensembles, setting a lower bound to their length. Specifically, estimation of high-order statistics needed for better separation of adjacent blood vessels requires sufficiently long ensembles in order to achieve good SNR. In addition, longer ensembles reduce the temporal resolution but increase the probability of detecting flow inside small blood vessels with extremely low UCA concentrations. The kidney scan processed in this work included 150 frames compared to the 1000 frame long tumor scan. In color and power Doppler modes frame rates higher than 6 Hz are used in order to ensure smooth image update [3]. Even with ensembles of 1000 frames, this rate can be achieved in plane-wave imaging by processing partially overlapping time-series. Therefore, the proposed method enables dynamic imaging of flow. In [16] 5000 frames were used to produce the SOFI images. The results presented in this work suggest that good SNR and background rejection can be achieved using CEUS scans with shorter ensembles. With this time-span a balance between temporal resolution and improved spatial resolution may be achieved.

As the SNR in CEUS scans is lower than in regular ultrasound scans [8], the effect of background noise in classic mean envelope images may be dominant. This could be a major problem in scans of poorly perfused tumors where UCA concentrations are low. For example, the background level in the mean envelope image in Figs. 6 and 7 masks the small vessels inside the tumor. The background noise level limits the effective dynamic range of the resulting mean envelope images. In this work, correlations and cumulants of *in-vivo* CEUS signals with single frame time-lags were used to minimize the influence of the white thermal noise. These images have reduced background levels, producing clearer images of the vessels and improving the dynamic range from 6dB to more than 60dB for the investigated small vessels. This improvement in SNR is significant even if it is normalized by the order of the statistics (60dB/4 = 15dB).

The use of statistics with different time delays present interesting characteristics and possible advantages. For example, the mean Doppler frequency in a given location is typically calculated from the auto-correlation of short temporal ensembles measured in each pixel [28]. Time lags different from zero are used in this case to track the phase changes resulting from the movements of the scatterers and estimate their velocity. The second moment with zero time-lag is used for power Doppler measurements in cases where the amplitude of the Doppler signal is of interest and not its direction. Cross-correlations between signals measured at neighboring pixels could improve SNR if the additive noise components in the different pixels are independent [17]. Moreover, cross-correlations and cross-cumulants calculated from neighboring pixels may serve as a way to perform interpolation and refining the grid of the resulting image [17]. The use of cross-statistics and different time lags are possible topics for future research.

There are two main limitations to the proposed framework. First, the dynamic range of the resulting image is highly affected by the order of the calculated statistics. Using linear display and high-order statistics, small blood vessels with low UCA concentration may potentially be overshadowed by large dominant vessels. However, log compression is standardly used in the display of ultrasound scans and the dynamic range of the logarithmic display is easily adjusted to fit the order of the statistics as done in this work. Second, the theoretic resolution gain of the proposed method improves proportionally to $\sqrt{p}$, where $p$ is the order of the statistics used. However, a moderate resolution gain is achieved, due to limitations on the dynamic range of the produced image and the number of acquired frames, limiting the order of the statistics used in practice. Since current methods yield either a low spatial resolution and excellent temporal resolution (classic CEUS) or outstanding spatial resolution and poor temporal resolution (ULM) as in [11], [12], [13], [14], a technique that presents a combination of improved spatial and temporal resolution like the one presented in this work is of great interest. We emphasize that currently the enhanced resolution presented in this work does not achieve the resolution gain of super-localization methods. However, in contrast to the proposed framework, super-localization microscopy methods were not reported to produce a clear image of the vasculature during the limited time intervals we consider. Provided that a separation between vessels of interest is attained using the proposed framework, deconvolution methods may be utilized to further improve the localization of the vessels in post-processing.

## VI. CONCLUSION

The methods studied in this work mitigate the trade-offs between spatial and temporal resolution in contrast-enhanced ultrasound imaging: tens or hundreds of milliseconds are needed in order to produce these estimations compared to up to a few minutes using ultrasound super localization microscopy methods ([11], [12], [13], [14]). With this short acquisition time, improved spatial resolution and background removal were presented. Although the gain in spatial resolution is modest compared to those achieved using super-localization methods, the short acquisition time makes the proposed approach clinically applicable. Improvement in temporal resolution of super-resolution ultrasound may enable the use of this technique to enhance cancer and neuroscience research where blood-flow dynamics are of great importance.